\documentstyle[11pt,paspconf,epsf]{article}
\def\etal{{et~al.~}}
\def\Lya{{\rm Ly}\kern 0.1em$\alpha$}
\def\HI{{\rm H}\kern 0.1em{\sc i}}
\def\Mg{{\rm Mg}\kern 0.1em{\sc ii}}
\def\MgI{{\rm Mg}\kern 0.1em{\sc i}}
\def\Fe{{\rm Fe}\kern 0.1em{\sc ii}}
\def\Ca{{\rm Ca}\kern 0.1em{\sc ii}}
\def\MgII{{\rm Mg}\kern 0.1em{\sc ii}~$\lambda\lambda2976, 2803$}
\def\C{{\rm C}\kern 0.1em{\sc iv}}
\def\NV{{\rm N}\kern 0.1em{\sc v}}
\def\CIV{C\kern 0.1em{\sc iv}~$\lambda\lambda1548, 1550$}
\def\kms{\hbox{km~s$^{-1}$}}
\def\cm2{\hbox{cm$^{-2}$}}

\begin{document}

\title{QSO Absorption Line Systems as a Probe
       of Galaxies Like the Milky Way\altaffilmark{1}}

\author{Jane C. Charlton\altaffilmark{2}}
\affil{Astronomy and Astrophysics Department, Pennsylvania State University,
       University Park, PA 16802}

\author{Christopher W. Churchill\altaffilmark{3}}
\affil{Board of Studies in Astronomy and Astrophysics,
       University of California, Santa Cruz, CA 95064}

\altaffiltext{1}{To appear in Galactic Chemodynamics 4: The History of 
the Milky Way and its Satellite System, eds. A. Burkert, D. Hartmann, 
and S. Majewski (PASP Conference Series)}
\altaffiltext{2}{Center for Gravitational Physics and Geometry,
             Pennsylvania State University}
\altaffiltext{3}{Visiting Astronomer, Astronomy and Astrophysics Department,
             Pennsylvania State University}

\begin{abstract}
Quasar absorption lines provide detailed information on
the chemical, kinematic, and ionization conditions in
galaxies and their environments, and provide a means for
studying the evolution of these conditions back to
the epoch of the first quasars.  
Among the collection of absorbing structures along the lines 
of sight to quasars there is an evolutionary sequence of 
galaxies that represent predecessors of the Milky Way and
provide a direct view of its history.
Absorption spectra of lines of sight through the Milky Way and
through nearby galaxies reveal a variety of chemical species,
ionization conditions, and kinematic substructures.  These absorption
profiles are produced by low density gas distributed in rotating
disks, high velocity halo clouds, satellite galaxies and their debris,
superbubbles, and other sub--galactic gaseous fragments.  
Guided by knowledge gained by studying nearby galaxies, we are
developing interpretations of the variety of observed absorption
signatures.  Images of $z\sim1$ galaxies responsible for {\Mg}
absorption also allow us to explore the statistical connections
between the galaxy properties and their gaseous content.  
Quasar absorption lines are fast becoming a powerful evolutionary probe
of gaseous conditions in the Milky Way.
\end{abstract}



\section{Introduction}

Quasar absorption line (QAL) studies have evolved past the stage of
simply counting the number of absorption lines due to a particular
ion as a function of equivalent width and redshift.  No longer do
we merely study disjoint classes of objects called {\Mg},
{\Lya}, and {\C} absorbers; we are now poised at the dawn of an era in
which we can explore the structures, metallicities, ionization
conditions, and kinematics within the individual galaxies that produce
the absorption lines.  Once the absorption signatures of Milky
Way--like galaxies are recognized, QALs can be used to trace the
detailed evolution of galactic gas from $z=5$ to $z=0$, the full
redshift range over which quasars have been observed.

Historically, QAL systems were divided into various categories based
upon how the samples were selected: 1) by {\Lya} (classified as
damped if the neutral column density $N_H > 10^{20.3}$~{\cm2},
Lyman limit if $10^{17.2} < N_H < 10^{20.3}$~{\cm2}, and forest if
$N_H < 10^{17.2}$~{\cm2}), 2) by {\Mg}, and 3) by {\C}.  Evolution in the
co--moving number per unit redshift\footnote{The co--moving number
is parameterized by $N(z) \propto (1+z)^{\gamma}$.  For no
evolution, $\gamma = 0.5$ in a $q_0=0.5$ universe and $\gamma = 1.0$
in a $q_0=0.0$ universe.} can be measured for each of these
``populations'' for a given detection threshold in equivalent width
(column density in the case of {\Lya}).  For $z>2$, the number of
{\Lya} forest systems is observed to rapidly decrease with time with 
$\gamma \sim 1.9$ (Bechtold 1994), and for $z<2$ is observed to slowly
decrease with $\gamma$ consistent with no evolution (Bahcall \etal
1996).   In the redshift interval $0.4 < z < 2.2$, the population
of {\Mg} absorbers ($W_0 > 0.3$~{\AA}) is consistent with no evolution,
though the subgroup of stronger absorbers ($W_0 > 1$~{\AA}) evolve
away with time very strongly ($\gamma \sim 2.2$).
The opposite trend, an increasing number with decreasing redshift, is
seen for {\C} systems ($W_0 > 0.15$~{\AA}) in the redshift interval
$1.2 < z < 3.7$ (Steidel 1990).  Overall, the statistical evolution of
these various populations is due to some combination of evolution in the 
metallicities, ionization conditions, and kinematic structuring of the
absorbing clouds.  A separation of these effects awaits detailed study
of the individual absorption line systems.

To better appreciate the inferences about the physical conditions of
the gas based upon absorption lines, we would like to know what type
of galaxy and what part of the galaxy is being probed by the QSO light
path. Specifically relevant to this proceedings is the question of
when we are looking through part of a Milky Way--like disk, high
velocity cloud, LMC--like satellite, and/or something like the LMC
Stream.  A good strategy
for sorting this out is to study nearby galaxies and the Milky
Way itself, to draw inferences on the kinematic, chemical, and spatial
distribution of the clouds from the absorption lines they produce,
and to ascertain what fraction of each absorber ``population'' is
presented by a galaxy of known type and morphology.  Deep imaging of
$z\leq1$ galaxies allows us to go further back in time and to
determine what luminosity and type of galaxy is associated with
particular QALs.  With the addition of theoretical modeling of the
expected absorption conditions arising from various processes
(eg.~winds, fountains, superbubbles, and infalling material), the
evolution of strengths, overall shapes, and subcomponent structures of
the absorption line profiles will allow us to infer a time sequence
in the Milky Way's history.

This article begins with a review of the insights that are likely to
be gained through absorption studies of the Milky Way and of nearby
galaxies.  The third section is a discussion of observations of {\Mg}
absorbing galaxies at $0.3 < z < 1.0$ that also focuses on studies
designed to discern the overall geometric cross sections of low
ionization galactic gas.  In the fourth section, we present HIRES
(Vogt \etal 1994) spectra of a number of {\Mg} absorbers (Churchill
1996a) and discuss what we hope to learn from the variety of
substructures that are observed.  The use of {\C} systems
at high $z$ in developing a picture of the history of the early Milky
Way galaxy is discussed in section five, as is the relevance of damped
{\Lya} systems for understanding the evolution of the Milky Way
disk.  We conclude with speculations on the physical nature of the 
objects that give rise to various types of QALs.

\section{Lines of Sight Through the Milky Way}

A sampling of lines of sight the Milky Way show a variety of structure
in the observed low and high ionization species.  The line of
sight to the low halo star HD~167756, which samples the inner galaxy
and disk, shows low ionization gas with velocities $-25 \leq v \leq
+30$~{\kms} (Cardelli, Sembach, \& Savage 1995).  Looking toward two
halo stars, Savage and Sembach (1994) find highly ionized gas with
{\C}/{\NV}~$>6$ toward the first and {\C}/{\NV}~$\sim$ 1--3 toward the
second.  The latter is consistent with the conditions predicted to
exist in superbubbles and galactic fountains.  The line of sight
toward 3C~273 passes through Galactic superbubbles which 
exhibit large {\C} and {\NV} column densities (Lu, Savage, \& Sembach
1994).  From the line of sight to the star HD~156359, which samples the
outer warp of the disk, a scale--height of 1~kpc for {\NV} and of
3--5~kpc for {\C} has been inferred.  This line of sight passes
through several spiral arms whose gas is seen to have a large velocity
dispersion due to turbulent mixing from supernova--related processes
(Sembach, Savage, \& Lu 1995).  The punch line here is that {\it even
looking through this specific galaxy at a particular stage in its
evolutionary history, we see substantial variation in the absorption
properties.}

Savage \etal (1993) compared the range of abundances of various ionic
species in the population of $z > 2$ damped {\Lya} absorbers to
those derived from lines of sight through the Milky Way.  They found
overlap between the two samples, with a trend toward larger line
strengths arising in the damped {\Lya} absorbers and a smaller number 
of strong lines from high ionization species in the Milky Way.  
Since all lines of sight used to study absorption through our
galaxy must pass through our region of the disk and through the lower
halo, the Milky Way sample is likely to be biased toward
low ionization conditions.
Nonetheless, it is interesting to speculate if redshift evolution in the
ionization conditions is occurring.  

How often does a line of sight through the Milky Way pass through a
high velocity cloud (HVC)?  The line of sight through the Milky Way toward
NGC~3783 passes through an HVC of $v\sim 240~${\kms} and a metallicity
of 15\% solar (Lu, Savage, \& Sembach 1994).  
In a recent survey, Murphy, Lockman, and Savage (1995) found an HVC covering
factor (for lines of sight passing through half the Milky Way) of
38\%.  However, Bowen, Blades and Pettini (1996a) surveyed for {\Mg}
HVCs (which should trace the Lyman limit {\HI}) and found that many
are consistent with co--rotation in the Galactic disk, and a covering
factor much smaller than 38\%.  Regardless, the population of Milky
Way HVCs is not sufficient to produce anywhere close to the unity
covering factor of {\Mg} absorbing gas that is inferred from studies 
of $z\leq1$ galaxies.  The high frequency with which galaxies
at intermediate redshifts are found to produce {\Mg} absorption is an
indication that either there is strong evolution of the HVC population
around galaxies, or other galactic components contribute significantly to
the cross--section of absorption.  It is interesting to note that
Bowen, Blades and Pettini find the LMC {\HI} ``double disk'' in {\Mg}
at impact parameter 7~kpc as well as Galactic {\Mg} along the line of
sight to Q0637--725.  

\section{The Geometry of {\Mg} Absorbing Gas}

In a GHRS survey of 17 lines of sight through nearby galaxies, Bowen,
Blades, and Pettini (1996b) find that lines of sight that pass
within 10~kpc of a galaxy center usually exhibit absorption while
lines of sight at impact $>30$~kpc do not.  
(Unfortunately, the paucity of known bright quasars at impact parameters
10--30~kpc from the centers of nearby galaxies prevents a robust
determination of local galaxy absorption covering factors.)  
There are exceptions to such a simple picture.  Lines of sight with
QSO--galaxy impact parameters $< 10$~kpc toward the early type
galaxies Leo~I and NGC~1380 do not exhibit absorption.  
On the other hand, the highly inclined disk toward galaxy G$1543+4856$
(smallest impact candidate) seems to be producing strong {\Mg}
absorption ($W_0 = 0.6$~{\AA}) at an impact parameter of 45~kpc.
Based upon their overall study, Bowen, Blades, and Pettini go on to
suggest that the environment of the absorbing galaxies has affected the
characteristics of the absorption, the strength of the lines, the
complexity of individual line components, and the ionization state of
the gas. 

For $z \leq 1$, Steidel, Dickinson, and Persson (1994, hereafter SDP)
used imaging and spectroscopy to identify the population of {\Mg}
absorbing galaxies that may represent predecessors to the
Milky Way.  In a sample of 58 absorbers in 51 QSO fields, they
identified galaxies that have the  same redshift as seen in {\Mg} 
absorption.  In a plot of the impact parameter $D$ versus the rest
frame $K$ luminosity, these 58 systems were found to define an
absorption cross--section ``boundary'', given by $D = 38h^{-1}
(L_K/L_K^*)^{0.15}$~kpc (Steidel 1995, hereafter S95).  Apparently,
nearly all galaxies produce {\Mg} absorption within $\sim 40$~kpc, and
very rarely is absorption seen at larger distances from a luminous ($>
0.05 L_K^*$) galaxy.  The most straight--forward interpretation is
that the gaseous galaxy halos have a relatively sharp edge within
which {\Mg} absorbing clouds are distributed with a nearly unity
covering factor.

In light of the above described studies of local galaxies, we
ask ``what is the contribution of galactic disks to the absorption
cross section''? It would appear that if {\Mg} absorption was produced
by disks, some galaxies with highly inclined disks would fail to
produce absorption at small impact parameters.  However, in these
inclined disks, the path length of the line of sight is increased due
to the orientation. For {\HI} disks that extend well beyond the
optical radius of the galaxy, the effect of passing further out in the
disk where the column density is smaller can be compensated by the
increased path length.  Realistically, a disk responsible for {\Mg}
absorption would be thick and consist of discrete absorbing clouds.
Even in the saturated regime, orientation can lead to an increased
equivalent width if the larger number of clouds along the path in an
inclined galaxy disk are spread over a larger range of velocities.  

The argument that {\Mg} absorption is often produced in the outer
disks of spiral galaxies is confirmed by 21 cm maps that show {\HI} at
radii of tens of kpc (Irwin 1995, Corbelli, Schneider, \& Salpeter
1989, van Gorkom \etal 1993).  Also, Bowen, Blades, and Pettini
(1995b) demonstrated that the radius for {\HI} absorption at
$N$({\HI})$=10^{20}$~{\cm2} increases with a larger power of the
luminosity $R = 16 (L/L_B^*)^{0.56}$~kpc than at the $10^{19}$~{\cm2}
contour level, for which $R = 23 (L/L_B^*)^{0.36}$~kpc.  If this trend
continues to lower $N$({\HI}), the disks of relatively low luminosity
spiral galaxies likely make a significant contribution to the {\Mg}
absorption cross section. Radio maps are not sensitive down to the
{\HI} column densities of $10^{17}$~{\cm2}, the level at which {\Mg}
is known to be associated, yet even the contour level of
$10^{19}$~{\cm2} shows that many galaxy disks will present a
significant cross section for {\Mg} absorption.  The shape of the
radial cutoff for {\Mg} disks depends on the physics of the photoionization
(Maloney 1993, Corbelli \& Salpeter 1993, Dove \& Shull 1994), so the
cross section of a ``population'' of {\Mg} disks is uncertain by about
a factor of two.

\begin{figure}[bh]
\plotfiddle{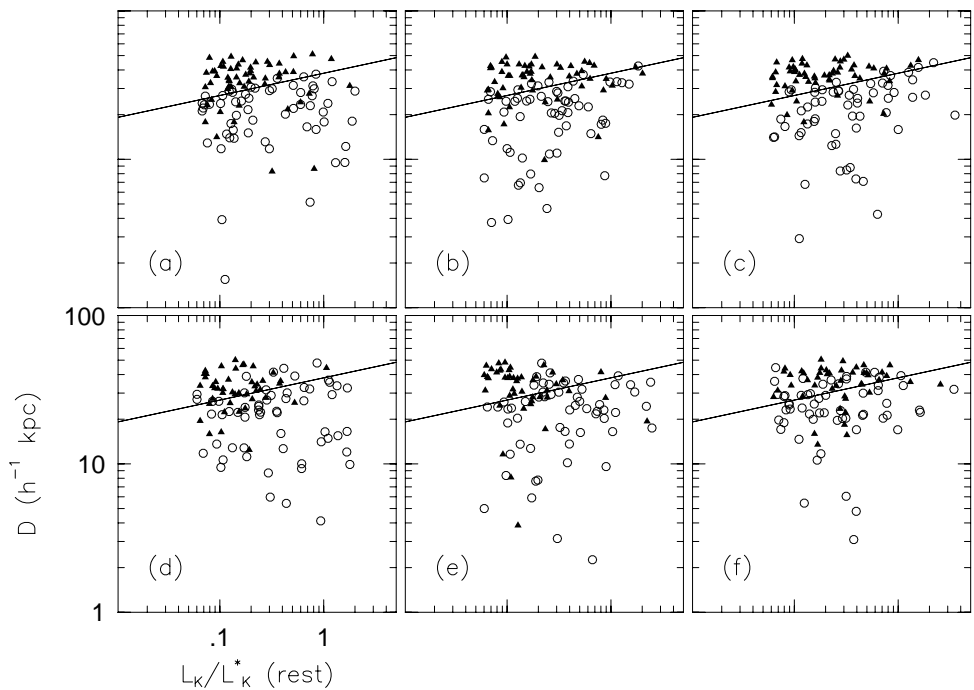}{3.25in}{0}{52.5}{52.5}{-200}{-35}
\caption{The distribution of impact parameter versus normalized
$K$ luminosity for three random realizations of a spherical
cloud model (upper panels) and of a disk model (lower panels).  
Open circles denote absorbing
galaxies and filled triangles denote non--absorbing galaxies.
These realizations sample the number of fields required to
produce 58 absorbing galaxies, as observed by SDP.  The model
then yields the number of non--absorbers for this same
number of fields, unbiased with respect to whether absorption
occurs in a field or not.  The solid line is the S95 best--fit
``boundary'' to the SDP data, $D=38h^{-1}(L_K/L_K^*)^{0.15}$~kpc.} 
\end{figure}

These ideas motivated us to design Monte--Carlo simulations that would
compare the expected absorption properties of populations of {\Mg}
clouds in spherical and in disk geometries (Charlton and Churchill
1996).  In Figure 1, we  illustrate results for absorbing and
non--absorbing galaxies in a simulated SDP study.  Both the spherical
and disk geometry models yield predicted numbers of non--absorbing
galaxies at small impact parameters and of absorbing galaxies at large
impact parameters than are larger than what is observed.  Spherical
cloud models with a unity covering factor are discrepant with the
observed distribution of {\Mg} equivalent widths, producing relatively
too many large values. We predict that spherical cloud models have
only a covering factor of 70--80\%, which is, in fact, similar to the
effective covering factor of the population of model disks.

\begin{figure}[bh]
\plotfiddle{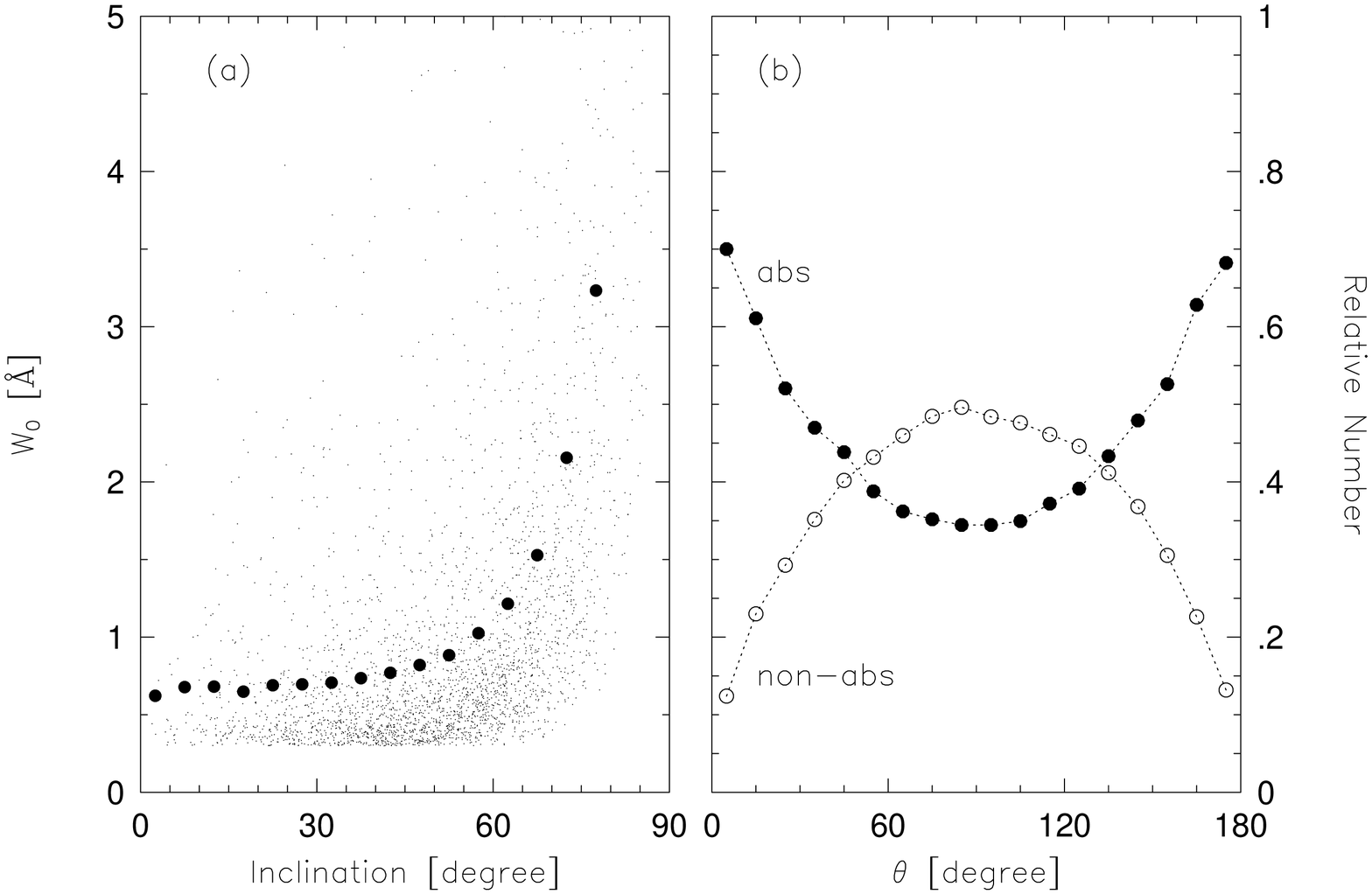}{3.25in}{0}{50.}{52.5}{-200}{-35}
\caption{Predictions from Monte--Carlo models 
for a randomly oriented population of absorbing disks.
--- (a) expected equivalent width of {\Mg} as a function of the
inclination.
--- (b) relative number of absorbers as a function of the position
angle of the QSO with respect to the projected inclined disk
major axis.}
\end{figure}

If galaxy disks play a significant role in defining the geometric
cross--section for {\Mg} absorption there should be a correlation
between the equivalent width of the absorber and the disk inclination
(due to the increased path length and resulting increased velocity
dispersion of the clouds in highly inclined disks).  A Monte--Carlo
model realization illustrating the expected trend is given in Figure
2a.  Hubble Space Telescope WFPC2 images will allow the orientations
and morphologies of the absorbing galaxies at $z<1$ to be determined.
With a sample of $\sim 20$ such images (being obtained by Steidel and
collaborators), it should be possible to utilize this test to
determine if disks make a significant contribution to the overall
cross--section for {\Mg} absorption.  A second prediction for the disk
population is illustrated in Figure 2b.  The cross section for
absorption is highest when the angle, $\theta$, subtended between the
QSO line of sight and the major axis of an inclined disk projected
on the sky, is zero or $\pi$.

We ``calibrated'' our simulated fields to account for the
selection and observational procedure employed by SDP.  First, we
considered that the observational sample was not 
unbiased in that it contained a
larger fraction of absorber fields than ``control'' fields and thus
was deficient in the total number of non--absorbing galaxies.  Second,
we considered that redshifts were not obtained (in the S95
results\footnote{The SDP survey is still in progress.  It is their
goal to obtain all redshifts within a well defined angular separation
from the QSO.}) for all galaxies at large distances from the quasar in
the absorber fields (i.e.~once the galaxy responsible for absorption
was identified the search was often stopped).  When the selection
procedures of S95 are applied to our model results we find little
distinction between either geometry, i.e.~for both only a few to
several non--absorbing galaxies at impact parameters less than
$38h^{-1}(L_K/L_K^*)^{0.15}$~kpc are observed in the model fields.
Discrepancy in these numbers are further reduced when possible
misidentifications of absorbing galaxies are incorporated into the
statistics.  We conclude that both the spherical and the disk
models are consistent with the current observations of the
population of {\Mg} absorbers.  In fact, disk gas, halo gas, and
satellite gas all surely play a role in shaping the {\Mg} absorption
profiles, and the overall geometric cross section of an absorbing
galaxy is unlikely to be the same at all column density levels.

\section{The Variety of {\Mg} Absorbers}

The idea that galaxy disks, clouds in galaxy halos, and satellite
galaxies all contribute to {\Mg} profiles is reinforced by the great
variety of substructure in Keck/HIRES spectra of {\Mg} absorbers at
6~{\kms} resolution.  A sampler of the kinematic substructure exhibited
by the ($\lambda 2796$) transition of the {\Mg} doublet is
presented in Figure 3.

These spectra are likely to be showing us absorption from satellites
of $L^{\ast}$ galaxies, rotating galaxy disks, multiple clumps in galaxy
halos, groups of galaxies, superbubbles, infall of gaseous fragments,
and/or outflow.  Using the observational clues from the Milky Way and
nearby galaxies in
tandem with theoretical models of the processes (such as fountains,
winds, and tidal stripping) that give rise to the absorption, we hope
to learn to recognize what galactic substructures are giving rise to
these complex profiles.  

As described in \S 2, the Milky Way itself exhibits large
differences in absorption along different lines of sight. Undoubtedly,
very different profiles in Fig.~3 could represent lines of sight
through different regions of galaxies that were similar to each other
in their gross properties.  Images of the absorbing galaxies (allowing
determination of impact parameters, luminosities, morphologies, and
orientations) will be a key to interpretation of the absorption
signatures (Churchill, Steidel, \& Vogt 1996, Churchill 1996b).  If
the absorption profiles can be decoded, such that the internal
dynamics, chemical compositions, and ionization states of galactic gas
can be inferred, the ultimate contribution of QAL studies would be
realized.  The processes by which galaxies of various types form and
evolve could be studied in detail back to the highest redshifts.

\begin{figure}[th]
\plotfiddle{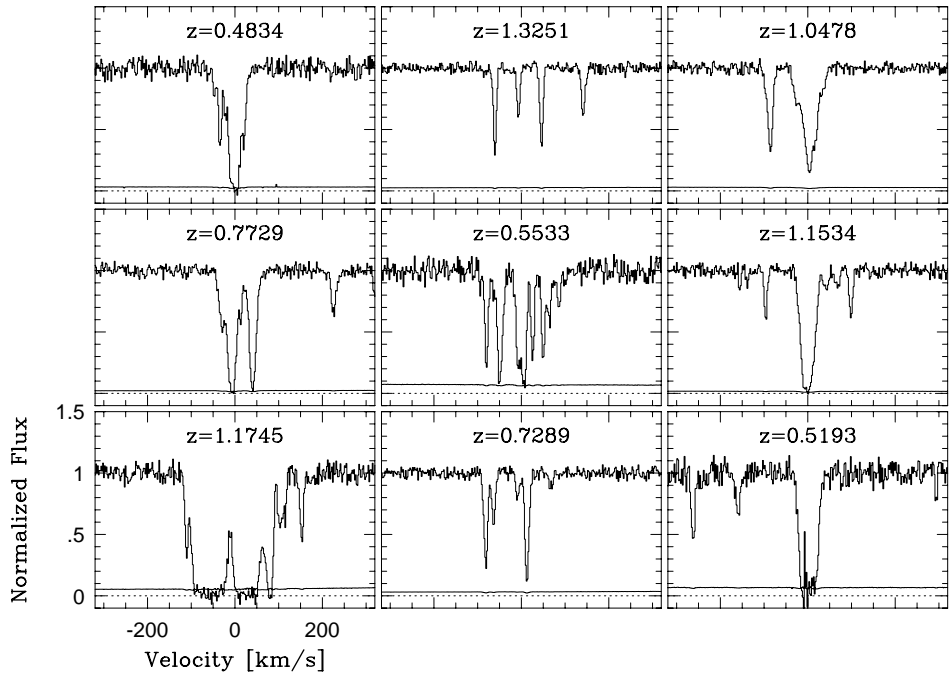}{3.25in}{0}{52.5}{52.5}{-200}{-35}
\caption{The HIRES {\Mg} profiles ($R=45000$) of
a sample of {\Mg} absorbers at $0.4 < z < 1.5$.
Absorption profiles have also been obtained for the low
ions {\MgI}, {\Fe}, and {\Ca}.  The full sample
will be presented elsewhere (Churchill 1996a).}
\end{figure}

\section{Clues from {\C} and Damped {\Lya} for Milky Way Evolution}

Here, we briefly comment on other examples of the power of QALs in
understanding the dynamical, chemical, and ionization evolution of the
gaseous components of the Milky Way.

If the Milky Way is any indication, {\C} and other high ions can be
used to trace processes such as galactic fountains and high velocity
clouds.  For example, the number of strong {\C} doublets decreases
with increasing redshift, but the weaker lines increase in number back
to at least a redshift of $z=3.7$ (Steidel 1990).  This is interpreted
as an increase in abundance of carbon by a factor of three in a typical
galaxy halo in just one billion years.  This interpretation, however,
is not definitive in light of the recent detection of a strong {\Mg}
absorption system at $z>4$ (Elston \etal 1996).  Since {\C} is
observable from the ground at these high redshifts, the discovery of
bright high redshift QSOs may facilitate a study of the buildup of
the first metals.

Wolfe (1995, references therein) has continued to suggest that damped
{\Lya} systems are the high redshift predecessors of Milky Way--like
disks.  This view is supported by the observation that the {\HI}
contribution to the total mass density of the Universe from the
population of damped {\Lya} absorbers decreases with time, consistent
with the expectation that gas is being converted into stars (Wolfe
\etal 1995).  Will further study of damped {\Lya} absorbers and their
metal content be key to understanding the past evolution of the
Milky Way disk?  Unfortunately, there are several serious
complications that must first be considered: 1) The several nearby
($z<1$) damped {\Lya} systems are not ordinary spirals, but are
instead found to be low surface brightness galaxies and dwarfs
(Steidel, private communication). 2) Dust obscuration could
seriously bias samples by eliminating from quasar catalogs those that
are obscured by the largest column density damped {\Lya} systems (Fall
\& Pei 1995).  3) Gravitational lensing could bias samples, adding to
the numbers of the largest column density absorbers (Bartelmann \& Loeb 1996).
4) It is quite rare to observe damped systems with values of
$N$({\HI}) as large as the stellar surface density in the centers of
galaxy disks at present (Salpeter 1996).  These facts suggest that either
infall plays a key on--going role in the evolution of spiral disks, or
that many of the high redshift damped {\Lya} systems are also low
surface brightness galaxies.  These complications will substantially
delay definitive conclusions for the Milky Way disk evolution based on
the study of damped {\Lya} absorbers.

\section{Speculations on the Origins of QALs}

We conclude with general hypotheses about the types of gaseous 
systems that give rise to the various populations of QALs:

\begin{enumerate}
\item 
Damped {\Lya} systems may mostly consist of the thick predecessors of
galaxy disks, but there is likely to be a non--negligible contribution
from lower surface brightness disks that do not evolve into Milky
Way--like galaxies.
\item 
It is likely that
Lyman limit/{\Mg} systems usually have an absorption contribution from
the extended disk of a spiral galaxy.  The multiple components of these
systems, extending over a range of $\sim$ few 100~{\kms}, provide
information about the environment of the dominant absorbing galaxy.
We should occasionally see galaxy pairs and groups in {\Mg}
absorption. This will provide a way of observing the evolution of the
environments and satellite systems of galaxies back to early times.
\item 
{\C} systems trace a more diffuse, more ionized component that may 
allow study of high velocity clouds and halo structure.  The
substructure observed in {\C} systems may relate to the infall
mechanisms and merger processes during the formation of galaxy halos.
\item 
{\Lya} forest systems are flattened and filamentary structures
(extending over hundreds of kpc) that form almost everywhere at high
redshifts and persist out to large distances around galaxies and
galaxy groups at present.
\end{enumerate}

\acknowledgments

This work was supported by NASA grant NAGW-3571 at the
Pennsylvania State University and by a California
Space Institute grant issued to Steven S. Vogt.  The ideas presented
in this proceedings took shape through conversations with many
colleagues, but special thanks are due to D. Bowen, M. Dickinson,
K. Lanzetta, C. Steidel for their insights, and especially to S. Vogt
whose HIRES opened the floodgate for detailed studies of the evolution
of the gaseous content of galaxies over the history of the Universe
via QSO absorption lines.

\end{document}